\documentclass[amsmath,amssymb,showpacs,rotating,twocolumn]{revtex4}

\usepackage{graphicx}
\usepackage{epsfig}
\usepackage{dcolumn}
\usepackage{times,mathptm,amsmath,amssymb,graphicx,here,fancyhdr}

\begin{document}


\title{Mirage phenomena in superconducting quantum corrals}

\title{Mirage phenomena in superconducting quantum corrals}
\author{Markus~Schmid and Arno~P.~Kampf}
     \address{Theoretische Physik III, Elektronische Korrelationen und
      Magnetismus, Institut f\"ur Physik, Universit\"at Augsburg, 86135
      Augsburg, Germany}
\date{\today}
	       
\begin{abstract}
We investigate the local density of states and the 
order parameter structure inside an elliptic quantum corral on 
surfaces of isotropic and anisotropic superconductors. 
The Bogoliubov-de Gennes equations are solved 
in the presence of non-magnetic and magnetic impurities. 
We observe and discuss a variety of mirage and anti-mirage 
phenomena, which specifically reflect the nature of the 
superconducting pairing state. 
\end{abstract}



\maketitle                

\section{Introduction}
Understanding impurity effects was one of the early, classical tasks for the 
theory of superconductivity \cite{Abrikosov63,Schrieffer64}. The Anderson 
theorem explained why non-magnetic impurity scatterers do not destroy 
conventional, isotropic $s$-wave superconductivity \cite{Anderson59}. For 
unconventional superconductors, however, both non-magnetic and magnetic 
impurities are pairbreaking and the sensitivity to impurities and the 
concomitant suppression of superconductivity may even serve as a fingerprint 
of an unconventional pairing state. It took three decades after the 
development of the BCS theory of superconductivity, before experimental studies 
of impurity effects on microscopic length scales became possible -- mostly due 
to the advance of scanning tunneling microscopy (STM). The need for 
experimental probes with a high spatial resolution arose from the discovery of 
a continuously growing number of unconventional singlet and even triplet 
superconductors; ruthenates and high-temperature cuprate superconductors 
belong to the most prominent examples. 

During the last decade it has furthermore become possible to manipulate 
surface structures on the atomic level. In particular the achievements in the 
research group of D. Eigler led to the design of closed atomic arrangements of 
adatoms on metallic surfaces \cite{Crommie93,Heller94}. In these quantum 
corrals with circular or elliptical shape electrons are confined to well 
defined geometries and STM techniques were applied to explore the structure of 
quantum mechanical wavefunctions and their interference patterns 
\cite{Heller94,Kliewer00}. Quantum mirage phenomena are observed, when an 
additional impurity atom is placed at one focus of an elliptic corral 
\cite{Manoharan00}. 
The terminology of quantum mirages is hereby defined by the observation that 
mirror images of the local density of states pattern around the impurity at 
one focus point appear also at the second impurity free focus. 
In subsequent studies even Kondo resonances could be detected at the 
impurity-free focus point, if a magnetic impurity adatom was placed in the 
other focus \cite{Aligia05}.

Similar mirage phenomena are expected for elliptic corrals on superconducting 
surfaces \cite{Morr}. The local suppression of the superconducting order parameter around 
an impurity at a selected point in the corral should be transferred to image 
structures due to standing wave patterns extending over the entire 
area of the elliptic corral. Superconducting quantum corrals therefore provide 
unique systems to study the combined effects of impurity induced fingerprint 
structures of the pairing state and the mirage phenomena of interfering 
quantum mechanical waves. With the assumption that a superconducting state is 
established in the bulk and on the surface of the substrate material we 
discuss in this paper the expected phenomena within the framework of BCS 
theory and the Bogoliubov-de Gennes equations.

\section{The model}
We consider the following mean-field Hamiltonian for an isotropic and an 
anisotropic singlet superconductor, respectively, with magnetic or 
non-magnetic impurities
\begin{eqnarray}
H = \sum_{\sigma} \int {\rm d}^2{\bf r}\; \psi^{\dagger}_{\sigma}(\mathbf{r}) 
H_0 \psi^{}_{\sigma}(\mathbf{r}) + H_{s/d} + H_{imp} 
\label{meanfieldH}
\end{eqnarray}
with $H_0 =(-\frac{\hbar^2}{2m}\nabla^2 - \mu) + V(\mathbf{r})$. 
$\mu$ is the chemical potential, and $V(\mathbf{r})$ is 
a hard-wall potential, which confines the electrons to the interior of an 
elliptic corral. The Schr\"odinger equation for particles in an elliptic 
geometry was previously solved analytically; the resulting eigenfunctions are 
a combination of Mathieu functions $ce_r$, $se_r$ and modified 
Mathieu functions $Ce_r$, $Se_r$ \cite{Schmid,Porras}: 
\begin{eqnarray}
\varphi^c_{r,n_c}(\theta, \eta) &=& 
ce_r(\theta, k_n^c) \; Ce_r(\eta, k_n^c), 
\label{Math1} \\ 
\varphi^s_{r,n_s}(\theta, \eta) &=& 
se_r(\theta, k_n^s) \;Se_r(\eta, k_n^s), 
\label{Math2}
\end{eqnarray}
where $\theta$ and $\eta$ are elliptical coordinates. $r$, $n_{c(s)}$ 
enumerate the quantum numbers for the eigenstates. Their eigenenergies 
\cite{Schmid} 
\begin{eqnarray}
\epsilon = \frac{2 \hbar^2}{(ae)^2 m} k^{c/s}_{r,n}
\end{eqnarray}
are determined by the zeros of the modified Mathieu functions $k^{c/s}_{r,n}$. 
$a$ denotes the length of the semimajor axis 
and $e$ the eccentricity of the ellipse; $m$ is the electron mass. 

For an isotropic $s$-wave superconductor with an attractive contact 
interaction $H_s$ is given by
\begin{eqnarray}
H_s = - \int{\rm d}^2{\bf r} \; \big[\psi^{\dagger}_{\uparrow}(\mathbf{r}) 
\psi^{\dagger}_{\downarrow}(\mathbf{r}) \Delta(\mathbf{r}) + h.c. \big], 
\end{eqnarray}
where $\Delta(\mathbf{r})=g\langle\psi^{}_{\downarrow}(\mathbf{r})
\psi^{}_{\uparrow}(\mathbf{r})\rangle$ is the order parameter and $g>0$ the 
pairing interaction strength. Similarly, the mean-field Hamiltonian for a 
singlet superconductor with a spatially extended pairing interaction is 
described by
\begin{eqnarray}
H_d =-\int{\rm d}^2{\bf r} \int{\rm d}^2{\bf r}' \;\big[\psi^{\dagger}_{
\uparrow}(\mathbf{r})\psi^{\dagger}_{\downarrow}(\mathbf{r'}) 
\Delta(\mathbf{r, r'}) + h.c.\big],
\end{eqnarray}
where the order parameter $\Delta(\mathbf{r, r'})=g(\mathbf{r, r'})\langle 
\psi^{}_{\downarrow}(\mathbf{r'})\psi^{}_{\uparrow}(\mathbf{r})\rangle$ 
depends on both coordinates $\mathbf{r}$ and $\mathbf{r'}$ of the electrons 
forming the Cooper pair. Our ansatz for the pairing interaction is 
$g(\mathbf{r, r'})=g \;\delta(|\mathbf{r-r'}|-R)$ and assumes an attraction 
for electrons at a distance $R$; this lengthscale $R$ should be set by the 
size of a typical crystal lattice constant. Although $g({\bf r},{\bf r}')$ is 
an isotropic real space interaction, it gives rise to an anisotropic local 
order parameter as we discuss below. The impurity potentials for magnetic ($-$) 
and non-magnetic ($+$) impurities are described by 
\begin{eqnarray}
H_{imp}=\int{\rm d}^2{\bf r} \; U(\mathbf{r}) 
\big[\psi^{\dagger}_{\uparrow}(\mathbf{r}) \psi^{}_{\uparrow}(\mathbf{r}) 
\pm \psi^{\dagger}_{\downarrow}(\mathbf{r}) 
\psi^{}_{\downarrow}(\mathbf{r}) \big], 
\end{eqnarray}
where $U(\mathbf{r})=\sum_i^n U_0 \delta(\mathbf{r - r_i})$ with impurity 
positions $\mathbf{r_i}$. In this model for the magnetic scattering center the 
impurity spins are treated as classical spins, $S\gg 1$ 
\cite{Shiba,Schrieffer}, which is a reasonable assumption e.g. for Mn and Gd 
adatom impurities on a niobium surface \cite{Yazdani}.  

The mean-field Hamiltonian Eq. (\ref{meanfieldH}) is diagonalized by the 
transformation 
\begin{eqnarray}
\psi_{\sigma}(\mathbf{r}) &=& \sum_n \big[u_n(\mathbf{r}) \alpha_{n\sigma} 
+ sgn(\sigma) v_n(\mathbf{r}) \alpha_{n-\sigma}^{\dagger} \big],
\label{Bogtrans} \\
\alpha_{n\sigma}&=& \int{\rm d}^2{\bf r} \; \big[u_n(\mathbf{r})\psi_{\sigma}
(\mathbf{r})- sgn(\sigma) v_n(\mathbf{r}) \psi^{\dagger}_{-\sigma}(\mathbf{r}) 
\big], 
\end{eqnarray}
where $\alpha_{n\sigma}$ describes a superconducting quasiparticle with spin 
$\sigma$ and energy $E_n$. The ansatz (\ref{Bogtrans}) leads to the 
Bogoliubov-de Gennes equations 
\begin{eqnarray}
\big[H_0 + U(\mathbf{r})\big] u_n(\mathbf{r}) + \int{\rm d}^2{\bf r}' \; 
\Delta(\mathbf{r, r'}) v_n(\mathbf{r'}) = E_n u_n(\mathbf{r}), 
\label{bgd1} \\ 
\int{\rm d}^2{\bf r}' \; \Delta(\mathbf{r, r'})u_n(\mathbf{r'})-\big[H_0 \pm 
U(\mathbf{r})\big] v_n(\mathbf{r}) = E_n v_n(\mathbf{r}), 
\label{bgd2}
\end{eqnarray}
where the +/- sign in Eq. (\ref{bgd2}) corresponds to non-magnetic/magnetic impurities, 
respectively. These equations are simplified in the case of an isotropic 
superconductor, where the pairing-interaction is pointlike, and therefore 
$\Delta(\mathbf{r, r'})=\delta(\mathbf{r-r'}) \;\Delta(\mathbf{r})$. The order 
parameter $\Delta({\bf r})$ is determined self-consistently from the condition 
\begin{eqnarray}
\Delta(\mathbf{r})=g\sum_n u_n(\mathbf{r}) v_n(\mathbf{r}) \; 
\big[1 - 2 f(E_n) \big] \, ,
\label{ops}
\end{eqnarray}
where $f(E)$ denotes the Fermi function. For an anisotropic 
superconductor with a finite range pairing interaction we have instead
\begin{eqnarray}
\Delta(\mathbf{r, r'}) = g(\mathbf{r,r'}) \sum_n \big\{u_n(\mathbf{r}) 
v_n(\mathbf{r'}) \big[1 - f(E_n) \big] \label{opd} \\ \nonumber 
- u_n(\mathbf{r'}) v_n(\mathbf{r}) f(E_n) \big\}.
\end{eqnarray} 
In order to solve the Bogoliubov-de Gennes equations (\ref{bgd1}) and 
(\ref{bgd2}) we expand $u_n(\mathbf{r})$ and $v_n(\mathbf{r})$ in terms of 
free electron eigenfunctions inside the elliptic corral 
\cite{Schmid} 
\begin{eqnarray}
u_n(\mathbf{r}) &=& \sum_k u_{kn} \; \varphi_k(\mathbf{r}), \\
v_n(\mathbf{r}) &=& \sum_k v_{kn} \; \varphi_k(\mathbf{r}),
\end{eqnarray}
where $k$ enumerates all the eigenstates given in Eqs. 
(\ref{Math1}) and (\ref{Math2}). For the ground-state energies of the 
isotropic (Eq. (\ref{grs})) and the anisotropic (Eq. (\ref{grd})) 
superconductor we find
\begin{align}
E^s_g =& \;2 \sum_n \int{\rm d}^2{\bf r} \; \big[ v_n(\mathbf{r}) H_0 v_n(\mathbf{r})
- \Delta(\mathbf{r}) u_n(\mathbf{r}) v_n(\mathbf{r}) \big], 
\label{grs} \\ 
E^d_g =& \sum_n \int{\rm d}^2{\bf r} \; \bigg[2 v_n(\mathbf{r}) H_0 v_n(\mathbf{r}) \label{grd} 
\\ \nonumber
-& \int{\rm d}^2{\bf r}' \Delta(\mathbf{r, r'}) 
\big(u_n(\mathbf{r}) v_n(\mathbf{r'}) + 
u_n(\mathbf{r'}) v_n(\mathbf{r}) \big) \bigg].
\end{align}
The local density of states (LDOS) is obtained from 
\begin{eqnarray}
N(\mathbf{r}, \omega) = \sum_n \big[ u_n^2(\mathbf{r}) \delta(\omega - E_n) 
+ v_n^2(\mathbf{r}) \delta(\omega + E_n) \big]\, ,
\label{ldos}
\end{eqnarray}
and the total density of states (DOS) is $N(\omega)=\int{\rm d}^2{\bf r}\, 
N({\bf r},\omega)$.

\section{Results}
For the elliptic corral we choose its eccentricity $e=0.5$ and the length of 
the semi-major axis $a=150$ \AA; these two parameters determine the 
single-particle eigenenergies and -states. With an approximate electronic 
density of one electron per 6\AA $\times$6\AA\, square the total number of 
electrons in the corral is $N=1726$. With this choice the $863^{rd}$ 
eigenstate at the Fermi energy $\epsilon_F$ has a large probability density at the foci, 
which is a favorable situation for the occurrence of mirage phenomena. For an 
isotropic superconductor the mirage phenomenon is observed, whenever there is 
one of these non-zero probability density eigenstates lying inside the energy 
gap. For our choice of the pairing interaction strength 
$g=1.0\times10^{-3}\epsilon_F$ 
the energy gap $\Delta_0$ 
stretches over about 50 eigenstate energies, where several of these states 
have the required property. Hence, the existence of mirage phenomena for an 
isotropic superconducting surface does not depend sensitively on the position of 
the Fermi energy, in contrast to a metallic, normal conducting surface 
\cite{Schmid} and also in contrast to an anisotropic superconductor, 
as we discuss below. 

\begin{figure}[t]
\centering\epsfig{file=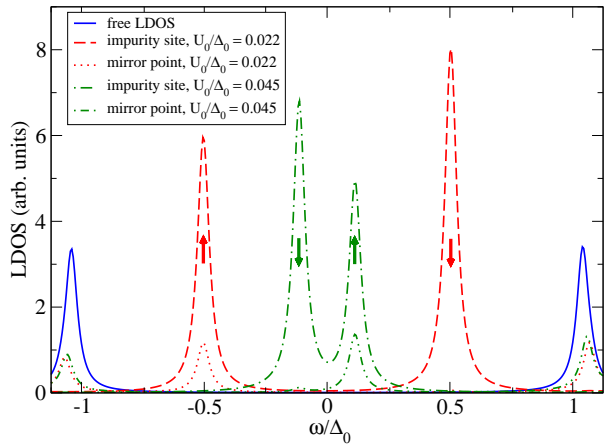, height=9.5cm, angle=-90}
\vspace{-1em}
\caption{Local density of states (LDOS) for an isotropic superconductor with a 
magnetic impurity at the right focus. Dashed and dash-dotted lines show the LDOS at the 
impurity site $(0.5a, 0)$, dotted and dash-double-dotted lines show the LDOS at the mirror point 
$(-0.5a, 0)$. The arrows ascribe the spin projection to the particle and the
hole component of the bound state.}
\label{fig1}
\end{figure}

\subsection{Isotropic $s$-wave superconductor}
\begin{figure}[t]
\centering\epsfig{file=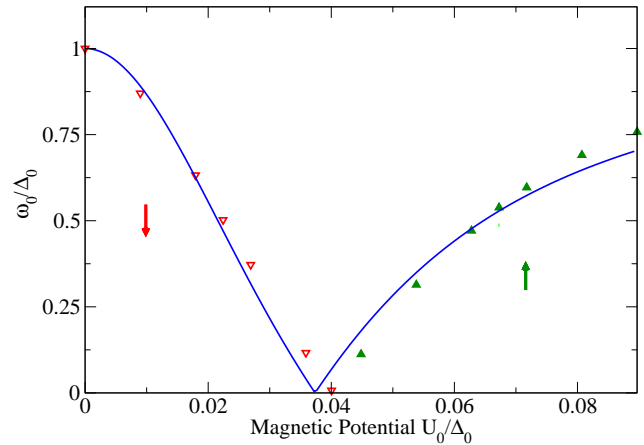, height=9.5cm, angle=-90}
\vspace{-1em}
\caption{Bound-state energy for an elliptic quantum corral (triangles) with a
magnetic impurity at the right focus $(0.5a, 0)$ compared with the
bound-state energy for a magnetic
impurity on an open surface of a $s$-wave superconductor (solid curve);
the arrows mark the
spin-direction of the particle component of the quasiparticle bound state.
Filled/unfilled triangles
correspond to a spin-down/spin-up quasiparticle, respectively;
$U\protect_0/\epsilon\protect_F \protect\cong 0.03 U\protect_0/\Delta\protect_0$.}
\label{fig2}
\end{figure}

We start our analysis with a corral on the surface of an isotropic 
$s$-wave superconductor. In the presence of non-magnetic impurities 
the system remains time-reversal invariant. These impurities do not break Cooper pairs, 
which are built from Kramers-degenerate electronic states \cite{Anderson59}. 
Consequently, no additional states are created in the renormalized 
energy gap. Magnetic 
impurities on the contrary, are pair-breaking defects, because the spin-up 
electron of the Cooper pair is repelled ($U_0>0$) and the spin-down electron 
is attracted by the impurity, leading to a quasiparticle bound state 
with a spin-down 
particle peak and a spin-up hole peak in the LDOS shown in Fig. \ref{fig1} 
(dashed and dotted curves), where the impurity is located at the right focus of 
the ellipse. 
The LDOS at the impurity site vanishes at the gap edge energies, 
while the LDOS at the mirror point (dotted line) is finite at these 
energies. The total spectral weight of both peaks, i.e. the spatially 
and energy (between the gap edge energies 
$-\Delta_0$ and $\Delta_0$) integrated 
LDOS is unity, 
characterizing one localized quasiparticle. Hence, the bound state 
corresponding to the dashed line is a local spin-down quasiparticle state
(see also Fig. \ref{fig2}, unfilled triangles), where the spin projection of 
the particle component is antiparallel and that of the hole component is 
parallel to the impurity spin. This is the same phenomenon known for 
the extended $s$-wave superconductor without confining boundaries 
\cite{Schrieffer}. The dash-dotted and dash-double-dotted curves in 
Fig. \ref{fig1} show the LDOS for a 
stronger impurity potential. The difference now is that the spins of the 
particle and the hole peak have changed, i.e. we find a spin-up quasiparticle 
(compare with Fig. \ref{fig2}, filled triangles). This change is due to the fact 
that both peaks cross the Fermi energy at a critical impurity potential 
strength $U_c$.

In the corral geometry we observe the two-peak structure in the LDOS not only 
at the impurity site e.g. at one focus point, but also -- with some 
attenuation -- at the impurity free focus (dotted and dash-double-dotted lines) \cite{Morr}. 
The mirage effect is only strong at one of the two peaks at the 
mirror point LDOS. 
There the particle peak carries very little spectral weight 
for $U_0<U_c$ (see Fig. \ref{fig1}, dotted line), and so does the hole peak for $U_0>U_c$. But with 
increasing $U_0$ the particle and hole peaks at the mirror point are 
both growing. 
We emphasize that 
the LDOS peaks at the mirror point are at the same resonance energies as in 
the LDOS at the impurity site, because they belong to the same bound state 
with energy $\omega_0$. This is the quantum mirage effect in the LDOS for an 
elliptic corral built on the surface of an isotropic $s$-wave superconductor. 
As expected, we find this effect to be very robust against a change in the 
Fermi energy and thus the electron density. Moreover this mirage effect 
remains almost unaffected by an additional perturbation in form of a second 
magnetic impurity placed anywhere else inside the corral. 

\begin{figure*}[t]
\epsfxsize=12cm\centerline{\epsfbox{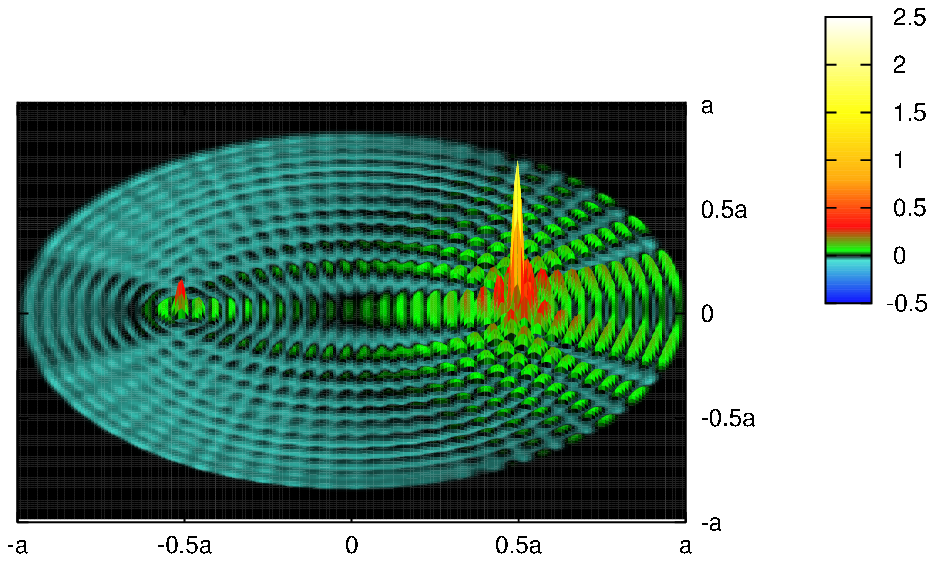}}
\caption{Difference plot of the local order parameter
$1\frac{\Delta^0(\mathbf{r})-\Delta^i(\mathbf{r})}{\Delta_0}$
in the elliptic corral; $\Delta^i(\Delta^0)$ is the order parameter in
the presence (absence) of the impurity;
the magnetic impurity is at the right focus $(0.5a, 0)$,
$U_0/\Delta_0 = 0.135$.}
\label{fig3}
\end{figure*}

Since the bound-state peaks at energies $\omega=\pm\omega_0$, split off from 
the gap edges, they move with increasing potential strength $U_0$ 
symmetrically towards the Fermi energy (see Fig. \ref{fig2}). 
$\omega_0$ therefore decreases 
until it reaches the chemical potential for a critical 
potential $U_c$, which depends on the impurity position 
${\bf r}_0$. 
At $U_0=U_c$ the bound state becomes a zero energy state, and the ground 
state of the superconductor becomes unstable. 
The existence of this critical point, which signals a 
ground-state level crossing transition, was first pointed out by Sakurai 
for $s$-wave superconductors with magnetic impurities \cite{Sakurai70}. 
At $U_c$ the system undergoes a first order phase transition from a spin zero 
to a spin $-\frac{1}{2}$ ground state \cite{Schrieffer,Sakurai70}. 
While the condensate 
ground state for $U_0 < U_c$ consists only of Cooper pairs, the new ground 
state for $U_0 > U_c$ contains an additional spin-down electron 
(or up-spin hole). The impurity induced 
bound state is now a spin-up quasiparticle 
(Fig. \ref{fig1}, dash-dotted and dash-double-dotted curves and Fig. \ref{fig2}, filled triangles). 
A further increase of $U_0$ in this regime ($U_0>U_c$) is accompanied 
by a spatial shift of the LDOS at the resonance energies away 
from the impurity 
position. This spatial redistribution of the spectral weight leads to 
decreasing heights of the bound-state peaks in the LDOS Fig. \ref{fig1} 
at the impurity position (right focus) 
and an accompanying increase of the peak heights at the impurity free focus. 
Note that the total spectral weight of the bound-state peaks 
always remains unity. 
Hence we observe a transition from a mirage-effect for
small $U_0$ to an anti-mirage-effect for large $U_0$.

We compare our self-consistent solutions of the Bogoliubov-de Gennes equation 
for a corral with the results for an open surface using the following 
formula for the bound-state energies \cite{Schrieffer}
\begin{eqnarray}
\omega_0 = \Delta_0 \frac{|1 - (\pi N_F)^2 U_0^2|}{1 + (\pi N_F)^2 U_0^2}, 
\label{bsenergies}
\end{eqnarray}
where $N_F$ is the normal state DOS at the Fermi energy. Eq. 
(\ref{bsenergies}) is obtained from the non-self-consistent T-matrix 
formalism, which proved to be sufficient and even quantitatively accurate 
for the description of magnetic 
impurity induced bound states in $s$-wave superconductors without confining 
walls \cite{Schrieffer}. In Fig. (\ref{fig2}) we use Eq. (\ref{bsenergies}) 
with $\pi N_F\rightarrow\overline{N_F}$ (solid curve), where $\overline{N_F}$ 
desribes the mean density of states averaged over an energy window 
$\epsilon_F\pm\Delta_0$ in the normal state. Fig. \ref{fig2} shows 
that the bound-state energies 
in the elliptic corral are well described by Eq. (19). 
Comparing the self-consistent 
and non-self-consistent results for our corral, we find indeed only small 
differences: neither the quasiparticle energies are shifted notably nor the 
heights of the quasiparticle peaks are modified significantly.

Another interesting quantity is the renormalized local order parameter 
$\Delta(\mathbf{r})$ given by Eq. (\ref{ops}). We observe a suppression of 
$\Delta(\mathbf{r})$ at the impurity site, where the lengthscale of the 
suppression and the wavelength of the order parameter oscillations is given 
approximately by the Fermi wavelength $\lambda_F=2\pi\hbar/\sqrt{2m\epsilon_F} 
\approx 15$ \AA, but seems to be independent of the size of the energy gap 
$\Delta_0$. Remarkably, the local order parameter at the impurity position 
changes discontinuously from a positive to a negative value, when 
$U_0$ 
crosses the critical potential strength $U_c$ \cite{Schrieffer, Flatte97}. 
For a non-magnetic 
impurity the reduction of $\Delta(\mathbf{r})$ in the vicinity of the impurity is due to 
the repulsive potential for both spin directions resulting in a reduced 
electronic density at and near the impurity site and therefore to a reduced 
local pairing amplitude. In Fig. \ref{fig3} a magnetic impurity is placed at 
the right focus point. All one-particle eigenstates with a large probability 
density at the right focus have a large probability 
density at the left focus as well. The difference plot of the order 
parameter Fig. \ref{fig3} strongly resembles the $863^{rd}$ eigenstate, 
implying that $\Delta(\mathbf{r})$ is suppressed not 
only at the impurity site (right focus), but almost symmetrically at the 
impurity free (left) focus, which we therefore call an order parameter mirage 
effect. 
We note that the order parameter is negative at the impurity site 
in Fig. \ref{fig3}.

If there are magnetic impurity spins in both foci, their antiparallel alignment 
leads to a lower ground-state energy Eq. (\ref{grs}). Moreover we observe that 
the order parameter suppression is about six times stronger for a parallel 
than for an antiparallel impurity spin alignment. 
A parallel alignment leads to a four-peak 
structure in the LDOS, 
because there are two similar states 
localized at each impurity, which are hybridizing and therefore 
splitting into bonding and antibonding states \cite{Flatte2000, Morr}. 
If the impurity spins are aligned antiparallel, we find only two peaks 
inside the energy gap. 
The intensity of the hybridization depends in 
general on the distance between the two impurities; 
for a corral geometry this dependence is quite complex.

\begin{figure*}[t]
\noindent
\begin{minipage}{0.47\textwidth}
\epsfxsize=12.8cm\centerline{\epsfbox{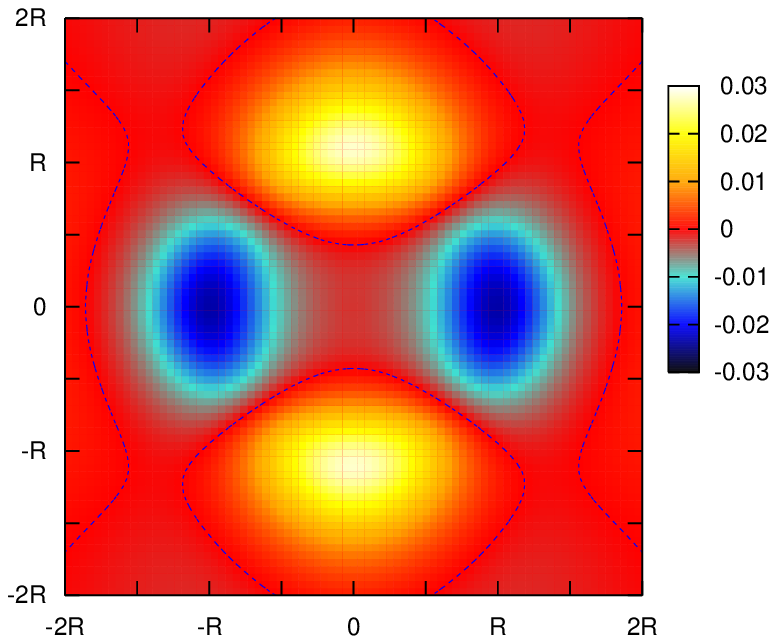}}
\end{minipage}
\hspace{-0.4cm}
\begin{minipage}{0.47\textwidth}
\epsfxsize=12.8cm\centerline{\epsfbox{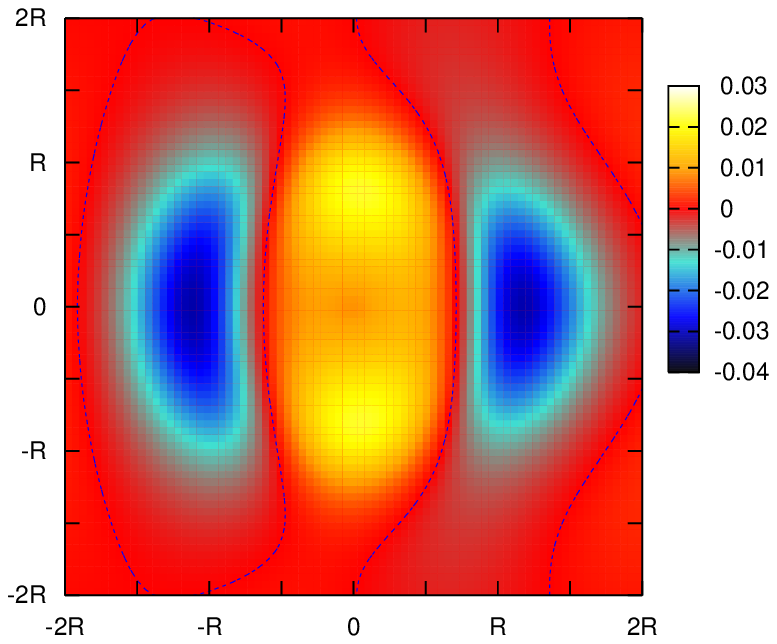}}
\end{minipage}
\vspace{-3em}
\caption{Local order parameter $\Delta(\mathbf{r,r+r'})$ in units of
$\epsilon_F$ for the anisotropic superconductor for a fixed ${\bf r}$ plotted
as a function of ${\bf r}'=(x',y')$ with $x', y'\in [-2R, 2R]$; the dashed
lines mark the order parameter nodes; $\mathbf{r} = (0, 0)$
in the left panel and $\mathbf{r} = (0.44a, 0)$ in the right panel.}
\label{op1}
\end{figure*}

\subsection{Anisotropic superconductor}

\begin{figure}[t]
\centering\epsfig{file=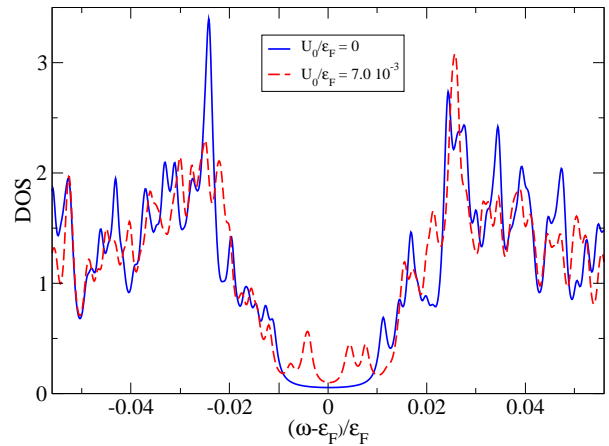, height=9.5cm, angle=-90}
\vspace{-1em}
\caption{Density of states (DOS) for
the anisotropic superconductor with a
non-magnetic impurity at $(0.44a, 0)$.}
\label{fig8}
\end{figure}

Now we assume the pairing interaction $g(\mathbf{r, r'})$ to be attractive in 
a distance of a typical crystal lattice constant $R\cong 0.04a=6$\AA. In this 
case the order parameter $\Delta(\mathbf{r, r'})$ depends on the position 
inside the corral as well as on the relative coordinate. Although the 
interaction is still isotropic, the order parameter is now anisotropic. In 
Fig. \ref{op1} (left panel) we show the order parameter in the center of the 
ellipse, where we identify a $d$-wave like structure with sign changes and 
nodes at angles near $45^\circ$ with respect to the semimajor axis. At the 
center the order parameter is least affected by the confining corral, but 
except for this point of highest symmetry, the structure of the local order 
parameter is distorted in a complex way. An 
example for a remarkably complicated symmetry and local structure of 
$\Delta(\mathbf{r, r'})$ is shown in the right panel of Fig. \ref{op1}. The spatially 
extended pairing interaction leads to a V-shape-resembling 
DOS (see Fig. \ref{fig8}, solid curve), with no sharp gap edges. Furthermore, 
the DOS has no particle-hole symmetry, because of the asymmetric distribution 
of eigenenergies for the corral eigenfunctions. 

\begin{figure*}[t]
\epsfxsize=12cm\centerline{\epsfbox{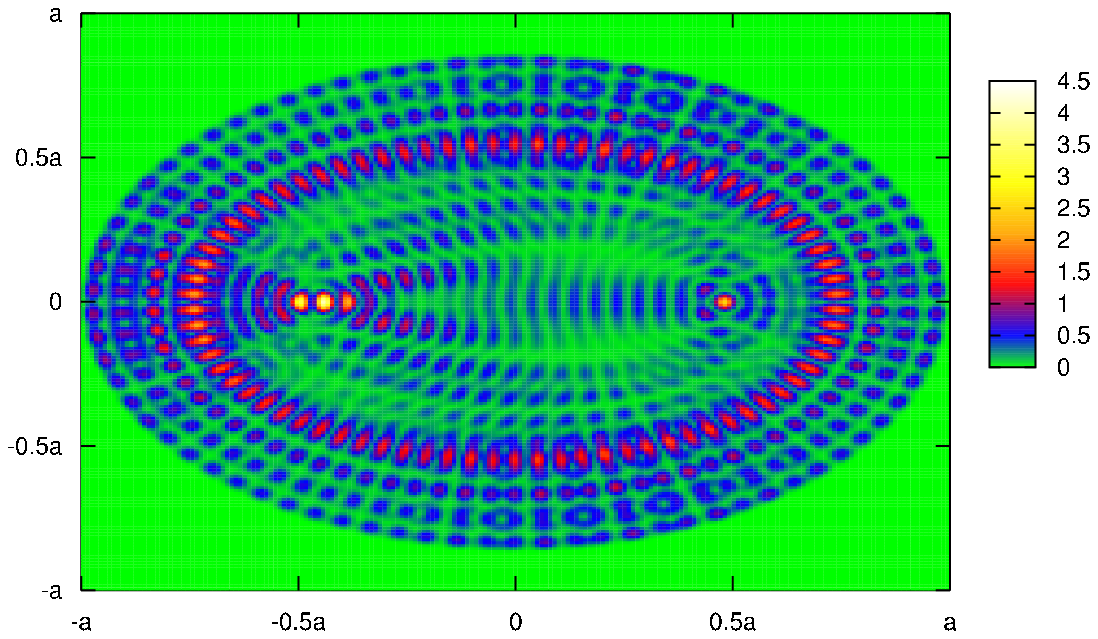}}
\caption{Local density of states (LDOS) for the anisotropic superconductor
with a non-magnetic impurity at $(0.44a, 0)$; $U_0/\epsilon_F=7.0
\times 10^{-3}$. The LDOS is shown at the positive resonance energy $\omega_L^+
=2.8\times 10^{-3}\epsilon_F$ (see text).}
\label{fig10}
\end{figure*}

\begin{figure*}
\epsfxsize=12cm\centerline{\epsfbox{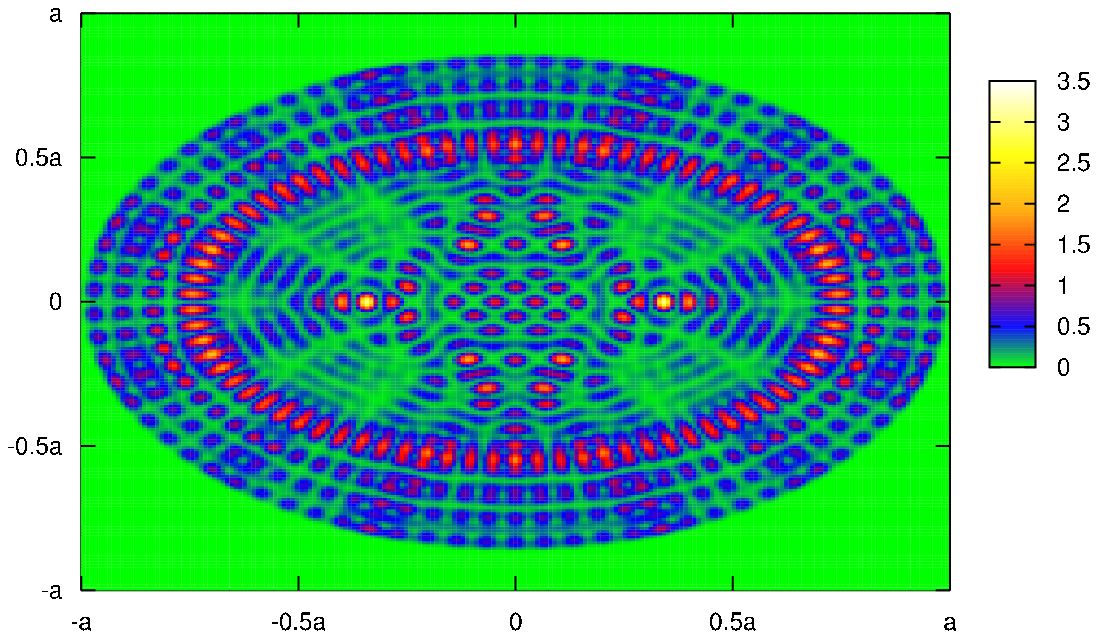}}
\caption{Local density of states (LDOS) with a non-magnetic impurity at the
center (0,0) of the ellipse; $U_0/\epsilon_F=14.0\times 10^{-3}$. The LDOS is
shown at the resonance energy $\omega_H^+= 7.6 \times 10^{-3}\epsilon_F$ (see
text).}
\label{fig9}
\noindent
\begin{minipage}{0.47\textwidth}
\epsfxsize=12.8cm\centerline{\epsfbox{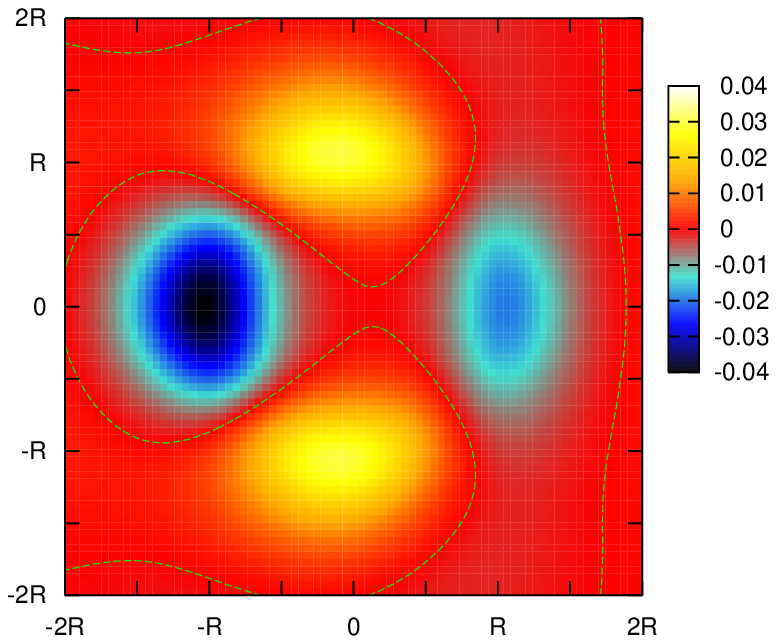}}
\end{minipage} 
\begin{minipage}{0.47\textwidth}
\epsfxsize=12.8cm\centerline{\epsfbox{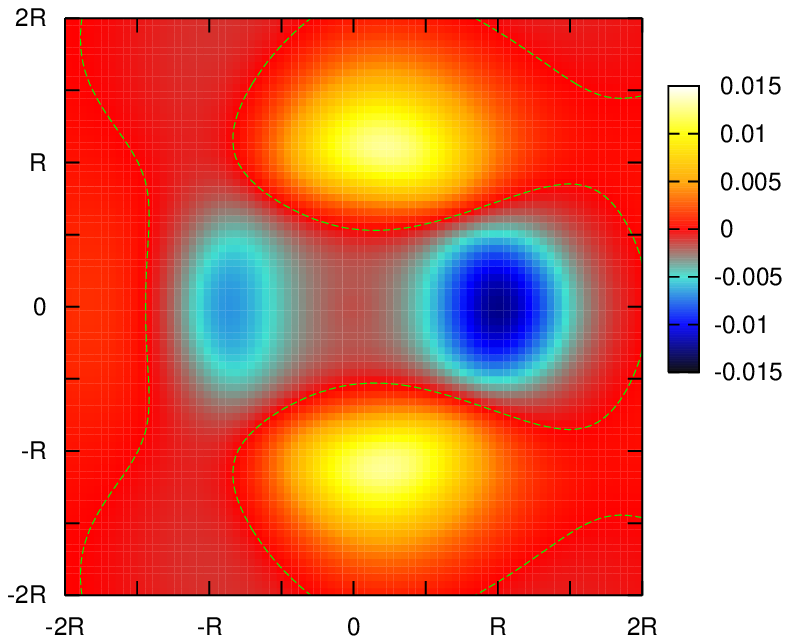}}
\end{minipage}
\vspace{-3em}
\caption{Local order parameter $\Delta(\mathbf{r,r+r'})$ for a fixed
$\mathbf{r}$ with a non-magnetic impurity at $(0.44a,0)$; $U_0/\epsilon_F=2.8
\times 10^{-3}$; $\mathbf{r} = (-0.44a, 0)$
in the left panel and $\mathbf{r}=(0.44a, 0)$ in the right panel. See also
the caption of Fig. 4.}
\label{op2}
\end{figure*}

While non-magnetic impurities do not give rise to localized states in a 
$s$-wave superconductor, they do have a pair-breaking effect in an anisotropic 
superconductor \cite{Gorkov}. For a non-magnetic impurity placed at different 
positions inside the corral we calculate the local order parameter, the DOS, 
and the LDOS. In the DOS we always find bound-state peaks, which move towards 
the Fermi energy as $U_0$ increases, but never reach $\epsilon_F$. Hence, we 
do not observe a zero-energy-peak in agreement with the results for an open 
surface in a particle-hole asymmetric situation \cite{Zhu,Schrieffer}. 
The DOS 
is modified over the entire energy range as shown in Fig. \ref{fig8}. 
For weak impurity potentials we observe a two-peak structure in the DOS near 
the Fermi energy, but after exceeding some critical potential 
$\tilde{U_c}$, which depends again on the impurity position $\mathbf{r_0}$, each 
of the two peaks starts to split into two, so that a four-peak structure 
emerges near $\omega=\epsilon_F$ (see Fig. \ref{fig8}). At the resonance energies 
for small $U_0$ (two-peak regime) we find, that the bound state has a large 
LDOS in the vicinity of the impurity. By increasing $U_0$ the original bound 
state acquires admixtures of the $862^{nd}$ and $866^{th}$ state 
\cite{Comment}. Above the critical potential ${\tilde U}_c$, we identify the 
$866^{th}$ state either at both outer or both inner peaks of the four-peak 
low-energy DOS, while the LDOS at the two other peaks 
consists of the small $U_0$ bound state mixed with a contribution of 
the $862^{nd}$ state. Hence the splitting of the bound state energy 
corresponds to a "demixing" of the contributing states, which form the bound state.
We observed this 
behavior at different impurity positions on the semimajor 
axis, but the occurrence of a four-peak structure nevertheless 
depends sensitively on the Fermi energy. Slight variations of the electronic 
densities can lead to two-peak structures in the DOS, irrespective of the 
impurity potential strength.

In order to show a few of a large variety of interesting and astonishing 
effects we calculate the LDOS in the presence of a non-magnetic impurity in 
particular at the right focus $(0.5a, 0)$, at $(0.44a, 0)$, and at the center 
(0,0). Interestingly, the important states ($862^{nd}$ and $866^{th}$) 
mentioned above are the only contributions to the bound state for an impurity 
in a focus point. These states have a zero LDOS on the entire semimajor axis, 
but several maxima about a lattice constant away from the focus. This bound 
state shows an almost perfect reflection symmetry with respect to the semiminor 
axis. Shifting the impurity a distance 1.5$R$ away from the right focus to 
$(0.44a, 0)$, we observe that the LDOS at the lower resonance energies 
$\omega_L^{\pm}=\pm 2.8\times10^{-3}\epsilon_F$ has a strong contribution of 
the $863^{rd}$ state and, as discussed before, a weaker contribution of the 
$862^{nd}$ state. Fig. \ref{fig10} shows the LDOS at the positive resonance 
energy $\omega_L^+$. While the LDOS at the impurity site itself is zero, we 
observe a maximum in the LDOS at the mirror point. The impurity 
induces asymmetrically spectral weight around both foci, whereas the height is larger 
at the left than at the right focus, which is close to the impurity.
Thus, we observe a surprising, strong anti-mirage effect in the LDOS at all 
resonance energies with a non-magnetic impurity at (0.44a, 0) independent of 
the potential strength $U_0$.

In Fig. \ref{fig9} we show the LDOS with an impurity located at the center of 
the ellipse at the energy of the higher positive resonance energy $\omega_H^+ 
=7.6 \times 10^{-3} \epsilon_F$. We identify the $862^{nd}$ eigenstate 
again, which determines the region near the boundary of the ellipse and 
consists of rings of peaks aligned along four 
ellipses. The structure of the LDOS in the vicinity 
of the impurity is in fact more influenced by the symmetry of the local 
order parameter than the geometry of the problem. 
E. g. in a range of a few lattice 
constants around the center there is no LDOS along the diagonal directions, 
and locally there is almost a fourfold symmetry as for the absolute value of 
the order parameter in Fig. \ref{op1} (left panel) \cite{Haas}.   

In the vicinity of the impurity site the order parameter is reduced strongly 
by a non-magnetic impurity. This becomes obvious by comparing the right panels 
of Figs. \ref{op1} and \ref{op2}, where the impurity is located at 
$(0.44a, 0)$. Moreover we observe here and quite generally a modificaton of 
the spatial structure and the symmetry of the local order parameter, 
which depends sensitively 
on the potential strength $U_0$. Interestingly the new, impurity 
induced spatial structure resembles more a $d$-wave-like 
symmetry than the original 
structure. Fig. \ref{op2} clearly indicates, that this $d$-wave-like order 
parameter structure is projected to the impurity free mirror point. Instead of 
a reduction, the order parameter is actually enhanced at the mirror point. As 
in the LDOS we therefore also observe an anti-mirage effect for the order 
parameter. The order parameter at $(0.44a, 0)$ is changed to 
a structure similar as in Fig. \ref{op2} (right panel), if an impurity is 
located at the center (!) of the ellipse. In general we find, that the 
magnitude of the order parameter is not only reduced over the entire area of 
the corral, but also modified with respect to its local spatial symmetry.

Regarding the connection between the spatial structure of the LDOS in the 
vicinity of the impurity and the order parameter, we encounter three 
different situations. (i) the LDOS resembles the order parameter 
in the absence of an 
impurity, (ii) the LDOS resembles the order parameter in 
the presence of an impurity, and 
(iii) there is no connection at all between the two. The latter is observed 
for example for a focus point impurity. We find the behavior (i) only for 
an impurity at the center of the corral, for large impurity potentials, 
and at positive resonance energies. For smaller impurity potentials we are
back to case (ii), where again the LDOS shows the structure of the order 
parameter with an impurity only at positive resonance energies. This second 
case seems to reflect the more generic situation.
Hence, the LDOS in the presence of a non-magnetic impurity inside the corral 
can directly reveal the structure and the local symmetry of 
the order parameter.

\section{Conclusion}
We have observed a rich variety of mirage- and anti-mirage-phenomena 
not only in the LDOS,
but also for the local order parameter structure. 
The induced patterns in both quantities in the vicinity of an impurity 
lead to mirror images in impurity-free regions, which are characteristic 
for both, the structure of the electronic wavefunctions in the elliptic geometry 
and the nature of the superconducting state. With the continuing advances of 
STM techniques and the already achieved capabilities for the design of atomic 
corral arrangements on metallic surfaces the realization of quantum corrals on 
superconducting surfaces indeed appears to be possible. From our model analysis 
we are led to expect intriguingly rich structures in the LDOS patterns, which 
by themselves  contain information about the nature of the superconducting state.

\begin{center}
{\small \bf ACKNOWLEDGEMENTS}
\end{center}
  One of the authors (A. P. K.) wishes to express his 
  special gratitude for the guidance and support during and ever after 
  his doctoral thesis work supervised by Bernhard M\"uhlschlegel. 
  This work was supported by the Deutsche Forschungsgemeinschaft
  through SFB 484.


\begin{thebibliography}{10}
\bibitem{Comment} We note that the bound states can be expanded in terms of 
the corral eigenfunctions $\varphi_k(\mathbf{r})$. Often we observe that only 
one or two of these states with energies near $\epsilon_F$ contribute 
significantly in the expansion of the bound state.
\end{thebibliography}
\end{document}